\documentclass[10pt,twocolumn]{IEEEtran}

\usepackage{cite}
\usepackage{graphicx}
\usepackage{epsfig}
\usepackage{psfrag}
\usepackage{subfigure}
\usepackage{url}
\usepackage{stfloats}
\usepackage{amsmath}
\usepackage{color}
\usepackage{amssymb}
\usepackage{algorithm}
\usepackage{algorithmic}
\usepackage{multirow}
\usepackage{array}
\usepackage{booktabs}
\usepackage{multirow}
\usepackage{threeparttable}

\usepackage{epstopdf}
\usepackage{epsfig}
\usepackage{psfrag}
\usepackage{subfigure}
\usepackage{bm}

\begin{document}
%
\title{Two-stage frequency recognition method based on correlated component analysis for SSVEP-based BCI}

\author{Yangsong Zhang$^{\#}$, Erwei Yin$^{\#}$, Fali Li, Yu Zhang, Toshihisa Tanaka, Qibin Zhao, Yan Cui, Peng Xu, Dezhong Yao, Daqing Guo*
\thanks{This work was supported in part by the National Natural Science Foundation of China under Grant 81571770, Grant 31771149, Grant 81401484, Grant 61522105, Grant 61527815, Grant 61703407, and Grant 61773129, in part by the Longshan academic talent research supporting program of SWUST under Grant 17LZX692, and in part by JSPS KAKENHI under Grant 17K00326. (Corresponding author: Daqing Guo. \# indicates that authors contributed equally to this work.)}
\thanks{Y. Zhang is with the Clinical Hospital of Chengdu Brain Science Institute, MOE Key Lab for Neuroinformation, University of Electronic Science and Technology of China, Chengdu 610054, China, and also with School of Computer Science and Technology, Southwest University of Science and Technology, Mianyang  621010, China. (e-mail: zhangysacademy@gmail.com)}
\thanks{E. Yin is with Unmanned Systems Research Center, National Institute of Defense Technology Innovation, Academy of Military Sciences China, Beijing, 100071 China.}
\thanks{Y. Zhang is with Department of Psychiatry and Behaviour Sciences, Stanford University, CA 94305 USA.}
\thanks{T. Tanaka is with   Department of Electronic and Information Engineering, Tokyo University of Agriculture and Technology, Koganei-shi, Tokyo 184-8588, Japan, also with Rhythm-Based Brain Information Processing Unit, RIKEN Center for Brain Science (CBS), Saitama 351-0198, Japan and Tensor Learning Unit, RIKEN Center for Advanced Intelligence Project (AIP), Tokyo 103-0027, Japan.}
\thanks{Q. Zhao is with Tensor Learning Unit, RIKEN Center for Advanced Intelligence Project (AIP), Tokyo 103-0027, Japan, and also with School of Automation, Guangdong University of Technology, Guangzhou, Guangdong, 510006, China.}
\thanks{F. Li, Y. Cui, P. Xu,  D. Yao and D. Guo are with the Clinical Hospital of Chengdu Brain Science Institute, MOE Key Lab for Neuroinformation, University of Electronic Science and Technology of China, Chengdu 610054, China. (e-mail: xupeng@uestc.edu.cn; dyao@uestc.edu.cn;  dqguo@uestc.edu.cn)}
}

\markboth{Manuscript for review}{}

\maketitle

\begin{abstract}
Canonical correlation analysis (CCA) is a state-of-the-art method for frequency recognition in steady-state visual evoked potential (SSVEP)-based brain-computer interface (BCI) systems. Various extended methods have been developed, and among such methods, a combination method of CCA and individual-template-based CCA (IT-CCA) has achieved the best performance. However, CCA requires the canonical vectors to be orthogonal, which may not be a reasonable assumption for EEG analysis. In the current study, we propose using the correlated component analysis (CORRCA) rather than CCA to implement frequency recognition. CORRCA can relax the constraint of canonical vectors in CCA, and generate the same projection vector for two multichannel EEG signals. Furthermore, we propose a two-stage method based on the basic CORRCA method (termed TSCORRCA). Evaluated on a benchmark dataset of thirty-five subjects, the experimental results demonstrate that CORRCA significantly outperformed CCA, and TSCORRCA obtained the best performance among the compared methods. This study demonstrates that CORRCA-based methods have great potential for implementing high-performance SSVEP-based BCI systems.

\end{abstract}

\begin{keywords}
Brain-computer interface, steady-state visual evoked potential, correlated component analysis, canonical correlation analysis, electroencephalogram.
\end{keywords}

%

\IEEEpeerreviewmaketitle

\section{Introduction}
A brain-computer interface (BCI) could provide an alternative communication pathway between the brain and a device. It can help severely paralyzed people  communicate or interact with their environment \cite{MBCILiyq2016,chaudhary2016brain}, and it assists healthy people, such as through autonomous driving \cite{Zander2017}. When designing a BCI system, noninvasive EEG is the most employed  brain imaging technique for extracting brain activity that codes the cognitive states and intentions of the user \cite{He2015}. The brain control signals  include event-related potential (ERP) \cite{JingJ2011,JINJ2014,Yin2v2016}, sensorimotor rhythm (SMR) \cite{Wuw2015,Zhangy2017JNS,FENG2018}, steady-state visual evoked potential (SSVEP) \cite{MFSC2012,Yin2015,Jiao2017}, hybrid BCI \cite{LiyqBME2013,Yin2014BME,Edelman2018}, and so forth.  SSVEP-based BCI has received increasing interest from researchers because it requires less training of the user and has a relatively high information transfer rate (ITR) \cite{Bakardjian2010, MSI2014,chen2015PNAS,chen2014,ZhangRMSI2016,HaiqiangSSVEP,Dean2017,zhangys2017MSI,Nakanishi2018}.

For SSVEP-based BCIs, developing an effective algorithm to recognize the SSVEP frequency with high accuracy and in a short time window (TW) is of considerable importantance for developing high-performance BCI applications. To date, various approaches have been proposed to recognize the SSVEP frequency. Among such methods, the canonical correlation analysis (CCA)-based recognition method  has been widely used to recognize targets due to its efficiency reported in the literature \cite{Lin2007,chen2014,Maye2017}. The standard CCA method, introduced by Lin \emph{et al.},  which uses sinusoidal signals as reference signals, was first proposed for SSVEP detection without calibration \cite{Lin2007}. However, the detection performance can be degraded by the interference from spontaneous EEG activities. Various extended methods have been proposed to incorporate individual EEG calibration data in CCA to improve the detection performance, such as  the cluster analysis of CCA coefficient (CACC) \cite{poryzala2014}, a phase-constrained CCA \cite{Pan2011}, individual-template-based CCA (IT-CCA) method \cite{Bin2011}, a combination method of CCA and IT-CCA \cite{nakanishi2015}, L1-regularized multiway CCA (L1-MCCA) \cite{L1MCCA}, the multiset CCA (MsetCCA) method \cite{MsetCCASSVEP, Suefusa2017}, and so forth. A comprehensive comparison among these methods was recently presented by  Nakanishi \emph{et al.}, and the results showed that the combination method based on the standard CCA and the IT-CCA achieved the highest performance \cite{nakanishi2015}.

CCA is a traditional technique for extracting linear combinations of data with maximal correlation \cite{Hotelling1936}, and it requires the canonical projection vectors (i.e., spatial filters) to be orthogonal. Unfortunately, this is not a meaningful constraint for EEG analysis. The spatial distributions are not expected to be orthogonal because they are determined by the current source distributions in space and the anatomy of the brain
\cite{Dmochowski2012}. Moreover, for two multichannel signals, CCA assigns two different projection vectors, thus doubling the number of free parameters and unnecessarily reducing the estimation accuracy. By dropping these constraints, a method named correlated components analysis (CORRCA) could be a promising alternative for designing frequency detection methods. CORRCA is derived from  maximizing the Pearson product moment correlation coefficient \cite{Pearson1896}.

In this study, we first introduce CORRCA as a standard method to implement frequency recognition, and then we propose a novel two-stage CORRCA method for frequency recognition. To evaluate the performance of the proposed methods, extensive comparisons are implemented among the standard CCA, the combination method of CCA and IT-CCA, standard CORRCA and TCORRCA using a benchmark dataset recorded from thirty-five healthy subjects. For all methods, the reference signals of each frequency are obtained by averaging the SSVEP data across multiple blocks. The experimental results indicate the promising potential of the proposed methods for accurately recognizing the SSVEP frequency in BCI applications.

The remainder of this paper is organized as follows. Section II presents the methods. Section III describes the experimental study. In Section IV, the experimental results on a benchmark dataset are reported. The discussion and conclusion are provided in the last two sections.

\section{Methods}
\subsection{Standard CCA}
CCA is a statistical method for measuring the underlying correlation between two sets of multidimensional variables, and it can find the weight vectors to maximize the correlation between the two variables \cite{Hotelling1936}. Given two multidimensional variables $\bm{X} \in \mathbb{R}^{m \times k} $ and  $\bm{Y}\in \mathbb{R}^{n \times k}$, CCA seeks a pair of weight vectors $\bm{w}\in \mathbb{R}^{m \times 1}$ and $\bm{v} \in \mathbb{R}^{n \times 1}$ such that the correlation between the resulting linear combinations  $\bm{x}=\bm{w}^{T}\bm{X}$ and $\bm{y}=\bm{v}^{T}\bm{Y}$ is maximized as:

\begin{equation}
\begin{aligned}
\rho&= \arg\max_{\bm{w},\bm{v}} \frac{\emph{E} \left[\bm{x}\bm{y}^{T}\right]}{\sqrt{\emph{E} \left[\bm{x}\bm{x}^{T}\right] \emph{E} \left[\bm{y}\bm{y}^{T}\right]}} \\
&=\arg\max_{\bm{w},\bm{v}}  \frac{\bm{w}^{T} \bm{X}\bm{Y}^{T} \bm{v}} {\sqrt{\bm{w}^{T}\bm{X}\bm{X}^{T}\bm{w}} \sqrt{\bm{v}^{T}\bm{Y}\bm{Y}^{T}\bm{v}}}
\,\,
\label{cca}
\end{aligned}
\end{equation}

Maximizing formula (\ref{cca}) can be achieved by solving a generalized eigenvalue problem. The maximum of $\rho$ with respect to $\bm{w}$ and $\bm{v}$ is the maximum canonical correlation.

CCA has been widely used for frequency recognition. In the standard CCA, the reference signals, i.e., $\bm{Y}_{i} \in \mathbb{R}^{2N_h \times N}$ ($i=1,2,\ldots,N_f$), are artificially created with the sine-cosine reference signals as follows \cite{Lin2007}:

\begin{equation}
\bm{Y_i}=\left(\begin{array}{c}
\sin(2\pi f_i t)\\
\cos(2\pi f_i t)\\
\ldots \\
\sin(2\pi N_h f_i t)\\
\cos(2\pi N_h f_i t)\\
\end{array}
\right), \; t = \frac{1}{Fs}, \frac{2}{Fs},\ldots, \frac{N}{Fs}
\end{equation}
where $N_h$ denotes the number of harmonics, $Fs$ is the sampling rate, and $N$ denotes the number of time samples.

With CCA, the maximum correlation coefficient $\rho_i$ can be computed between a test sample  $\bar{\bm{X}} \in \mathbb{R}^{C \times N}$ and each  $\bm{Y}_i$, $i=1,2,\ldots,N_f$, respectively. $C$ denotes the number of signal channels, and $N_f$ is the number of stimulus frequencies. Then, the frequency $f$ of the test sample was the frequency of the reference signals with the maximum correlation, as shown in formula (\ref{frule2}):

\begin{equation}
f_{\rm test}=\max_f \rho_{f}  ,  \; f = f_1, f_2,\ldots,f_{N_{f}} \,\, \label{frule2}
\end{equation}

\subsection{The combination method of CCA and IT-CCA}
In the IT-CCA method, the reference signals are individual templates obtained by averaging across multiple EEG trials from each subject \cite{Bin2011}. By replacing the artificial reference signals with the individual templates, the CCA process in this method is the same as standard CCA. The combination method of CCA and IT-CCA is an extended CCA-based method that combines the standard CCA and the IT-CCA approaches \cite{chen2015PNAS,nakanishi2015}. This method achieved the highest performance among the extended CCA methods. In this method, the feature of each frequency was not the maximum of $\rho$ in formula (\ref{cca}) but rather the correlation coefficient between the linear combination of a test sample $\bar{\bm{X}} \in \mathbb{R}^{C \times N}$  and an individual template $\bm{Z}_i \in \mathbb{R}^{C \times N} $ ($i=1,2,\ldots,N_f$) using CCA-based spatial filters. Specifically, the following three weight vectors were used as spatial filters: (i) $\bm{W}_{\bar{\bm{X}}}(\bar{\bm{X}} \bm{Z}_i)$  between the test sample $\bar{\bm{X}}$ and the individual template $\bm{Z}_i$, (ii) $\bm{W}_{\bar{\bm{X}}}(\bar{\bm{X}}\bm{Y}_i)$  between the test sample $\bar{\bm{X}}$ and sine-cosine reference signal $\bm{Y}_i$, and (iii) $\bm{W}_{\bar{\bm{X}}}(\bm{Z}_i \bm{Y}_i)$  between the individual template $\bm{Z}_i$ and sine-cosine reference signal $\bm{Y}_i$.  For the $i$-th template signal, a correlation vector $r_i$ was defined as follows \cite{chen2015PNAS}:

\begin{equation}
\begin{aligned}
r_i&=\left[\begin{array}{c}
r_i(1)\\
r_i(2)\\
r_i(3)\\
r_i(4)\\
r_i(5)\\
\end{array}
\right]\\
&=\left[\begin{array}{c}
\rho(\bar{\bm{X}}^{T}\bm{W}_{\bar{\bm{X}}}(\bar{\bm{X}}\bm{Y}_i), \bm{Y}^{T}\bm{W}_{y}(\bar{\bm{X}}\bm{Y}_i) )\\
\rho(\bar{\bm{X}}^{T}\bm{W}_{\bar{\bm{X}}}(\bar{\bm{X}}\bm{Z}_i), {\bm{Z}_i}^{T}\bm{W}_{\bar{\bm{X}}}(\bar{\bm{X}}\bm{Z}_i) )\\
\rho(\bar{\bm{X}}^{T}\bm{W}_{\bar{\bm{X}}}(\bar{\bm{X}}\bm{Y}_i),  {\bm{Z}_i}^{T}\bm{W}_{\bar{\bm{X}}}(\bar{\bm{X}}\bm{Y}_i) )\\
\rho(\bar{\bm{X}}^{T}\bm{W}_{\bar{\bm{X}}}(\bm{Z}_i \bm{Y}_i),  {\bm{Z}_i}^{T}\bm{W}_{\bar{\bm{X}}}(\bm{Z}_i \bm{Y}_i) )\\
\rho({\bm{Z}_i}^{T}\bm{W}_{\bar{\bm{X}}}(\bar{\bm{X}}\bm{Z}_i), {\bm{Z}_i}^{T}\bm{W}_{\bm{Z}_i}(\bar{\bm{X}}\bm{Z}_i))
\end{array}
\right]
 \,\,
\label{rcorr}
\end{aligned}
\end{equation}
where $\rho(\cdot,\cdot )$ indicates the computation of the correlation  between two signals. The number of harmonics was set to five to include the
fundamental and harmonic components of SSVEPs. The five correlation values described in  formula (\ref{rcorr}) were combined as the feature for target identification as follows:
\begin{equation}
\rho_{i}=\sum\limits_{k=1}^5 {\rm sign}(r_i(k)) \cdot (r_i(k))^2 \,\, \label{rcombination1} 
\end{equation}
where sign($\cdot$) was used to retain discriminative information from negative correlation coefficients. The target frequency of the test sample $\bar{\bm{X}}$ was then recognized by formula (\ref{frule2}).

\subsection{Standard CORRCA}
CORRCA is a technique that can produce the same weight vectors for two sets of multidimensional variables such that the linear components of two data are maximally correlated \cite{Dmochowski2012}. Its theoretical basis is to maximize the Pearson product moment correlation coefficient, and the weight vectors can be obtained by solving a generalized eigenvalue problem. CORRCA has been used to investigate cross-subject synchrony of neural processing \cite{CohenENEURO2016} and intersubject correlation in the evoked encephalographic responses \cite{dmochowski2014audience,Ki2016}.

Given two multidimensional variables $\bm{X}_1 \in \mathbb{R}^{C \times N} $ and  $\bm{X}_2 \in \mathbb{R}^{C \times N}$, where $C$ is the number of channels (i.e., electrodes) and $N$ is the number of time samples, CORRCA seeks to find a weight vector $\bm{w}\in \mathbb{R}^{C \times 1}$ such that the resulting linear combinations  $\bm{x}=\bm{w}^{T}\bm{X}_1$ and $\bm{y}=\bm{w}^{T}\bm{X}_2$  exhibit the maximum correlation.

\begin{equation}
\begin{aligned}
\hat{\rho}&=\arg\max_{\bm{w}}\frac{{\bm{x}}^{T}{\bm{y}}} {\left\| \bm{x} \right\| \ \left\| \bm{y} \right\|}\\
&=\arg\max_{\bm{w}} \frac{\bm{w}^{T} \bm{R}_{12} \bm{w}} {\sqrt{\bm{w}^{T}\bm{R}_{11}\bm{w}} \sqrt{\bm{w}^{T}\bm{R}_{22}\bm{w}}}\\
\end{aligned}
\,\, \label{corrca1}
\end{equation}
where $\hat{\rho}$ denotes the correlation coefficient. The sample covariance matrices are denoted as $\bm{R}_{ij}=\frac{1}{N}{\bm{X}_i}{\bm{X}_j}^{T}$, where $i,j=1,2$. Differentiating formula (\ref{corrca1}) with respect to $\bm{w}$ and setting to zero and assuming that $\bm{w}^{T}\bm{R}_{11}\bm{w} = \bm{w}^{T}\bm{R}_{22}\bm{w}$ leads to the following eigenvalue equation \cite{Dmochowski2012}:

\begin{equation}
(\bm{R}_{12}+\bm{R}_{21}) \bm{w} =\lambda (\bm{R}_{11} + \bm{R}_{22}) \bm{w}
\,\, \label{corrca2}
\end{equation}

The maximum of $\hat{\rho}$ corresponds to the principal eigenvector of  $ (\bm{R}_{11}+\bm{R}_{22})^{-1}(\bm{R}_{12}+\bm{R}_{21})$
that maximizes the correlation coefficient between $\bm{x}$ and $\bm{y}$.  Moreover, the second strongest correlation is obtained by projecting the data matrices onto the eigenvector corresponding to the second strongest eigenvalue and so on.

In this study, we propose a frequency recognition method based on CORRCA. To recognize the frequency of the SSVEPs with CORRCA, we can calculate the correlation coefficient $\hat{\rho}_i$ between a test sample $\bar{\bm{X}} \in \mathbb{R}^{C \times N}$  and an individual template $\bm{Z}_i \in \mathbb{R}^{C \times N} $ at each stimulus frequency, $i=1,2,\ldots,N_f$. The frequency ($f$) of the reference signal with the maximum correlation coefficient was selected as that of the test sample.

\begin{equation}
f_{\rm test}=\max_f \hat{\rho}_{f} ,  \; f = f_1, f_2,\ldots,f_{N_{f}} \,\, \label{frule}
\end{equation}

\subsection{Two-stage CORRCA}
Previous studies demonstrated that the spatial filters of different stimulus frequencies are similar to each other, and confirmed that integrating all the spatial filters could further improve the algorithm performance \cite{Nakanishi2018,zhangcorca2018}. Inspired by these studies, we propose a two-stage CORRCA method for frequency recognition based on standard CORRCA, which could utilize the spatial filters of all stimulus frequencies to yield more discriminative feature. In the first stage, we calculate the reference signals of each frequency by averaging the corresponding SSVEP data across multiple blocks with the individual training dataset and learn spatial filters for each frequency with the standard CORRCA. In the second stage, for each frequency,  we first calculate the correlation coefficients between a test sample and reference signals, and then we use all the spatial filters obtained in the first stage to calculate the correlation coefficients between a test sample and reference signals using the formula of the standard CORRCA. Then, all the correlation values are combined as the feature for target identification. The details of the computation are provided below.

Assume that $\bm{X}_{1,i},\bm{X}_{2,i},\ldots$ and $\bm{X}_{N_t,i}$ represent $N_t$ EEG trials of size $C \times N$ at the $i$-th stimulus frequency. Here, $N_t$ is the number of trials. Let $I_i=\{I_{i1},I_{i2}\}=\{(1,2),(1,3),\ldots,(N-1,N)\}$ denote the set of all $P=N \times (N-1)/2$  possible combinations of trial pairs at the $i$-th frequency, $i=1,2,\ldots,N_f$. Then, we can define two trial-aggregated data matrices as:
\begin{eqnarray}
\bar{\bm{X}}_{1,i}=[\bm{X}_{I_{11},i} \   \bm{X}_{I_{21},i} \  \ldots \  \bm{X}_{I_{P1},i}]\,\,.\\
\bar{\bm{X}}_{2,i}=[\bm{X}_{I_{12},i} \   \bm{X}_{I_{22},i} \  \ldots \  \bm{X}_{I_{P2},i}]\,\,.
\label{corrca3}
\end{eqnarray}

In the first stage, for the $i$-th stimulus frequency, we used the standard CORRCA of formula (\ref{corrca1}) to learn weight vectors $\bm{w}_i \in \mathbb{R}^{C \times 1}$ with $\bar{\bm{X}}_{1,i}$ and $\bar{\bm{X}}_{2,i}$, $i=1,2,\ldots,N_f$. In the second stage, with a test sample $\bar{\bm{X}} \in \mathbb{R}^{C \times N}$  and an individual template $\bm{Z}_i \in \mathbb{R}^{C \times N} $ ($i=1,2,\ldots,N_f$), we first calculated the correlation coefficient between  $\bar{\bm{X}}$  and $\bm{Z}_i$ with formulas (\ref{corrca1})-(\ref{corrca2}), denoted as $\beta_{i,0}$. Then, with the weight vectors $\bm{w}_k$ ($k=1,2,\ldots,N_f$), we further calculated the correlation coefficients $\beta_{i,k}$ between  $\bar{\bm{X}}$ and $\bm{Z}_i$  using the following formulas:

\begin{equation}
\begin{aligned}
\beta_{i,k}=\frac{\bm{w}_k^{T} \bm{R}_{12} \bm{w}_k} {\sqrt{\bm{w}_k^{T}\bm{R}_{11}\bm{w}_k} \sqrt{\bm{w}_k^{T}\bm{R}_{22}\bm{w}_k} },  \; k = 1, 2,\ldots,N_{f} \\
\end{aligned}
\label{betacombi}
\end{equation}

Here, the four sample covariance matrices are calculated as $\bm{R}_{11}=\frac{1}{N} \bar{\bm{X}} \bar{\bm{X}}^{T}$, $\bm{R}_{22}=\frac{1}{N}{\bm{Z}_i}{\bm{Z}_i}^{T}$, $\bm{R}_{12}=\frac{1}{N}\bar{\bm{X}}{\bm{Z}_i}^{T}$, and $\bm{R}_{21}=\frac{1}{N}{\bm{Z}_i}{\bar{\bm{X}}}^{T}$. For the $i$-th template signal, with  $N_f$  weight vectors, we can obtain a correlation vector $\beta_i$ defined as follows:
\begin{equation}
\beta_{i}=\left[\begin{array}{c}
\beta_{i,0}\\
\beta_{i,1}\\
\beta_{i,2}\\
\vdots\\
\beta_{i,N_f}\\
\end{array}
\right]
 \,\, \label{tcorrcc}
\end{equation}

These correlation values described in  formula (\ref{tcorrcc}) were further combined as the feature by the following formula:
\begin{equation}
\bar{\rho_{i}}=\sum\limits_{k=0}^{N_f} {\rm sign}(\beta_{i,k}) \cdot (\beta_{i,k})^2 \,\, \label{rcombination2}
\end{equation}
where ${\rm sign}( \cdot )$ was used to remain discriminative information from negative correlation coefficients as that in formula (\ref{rcombination1}) . Then, the frequency $f$ of the test sample $\bar{\bm{X}}$ was that of the template signals with the maximum correlation:

\begin{equation}
f_{\rm test}=\max_f \bar{\rho}_{f}  ,  \; f = f_1, f_2,\ldots,f_{N_{f}}  \,\, \label{frule1}
\end{equation}

\begin{figure*}[htbp]
\begin{center}
\includegraphics[width=0.95\textwidth]{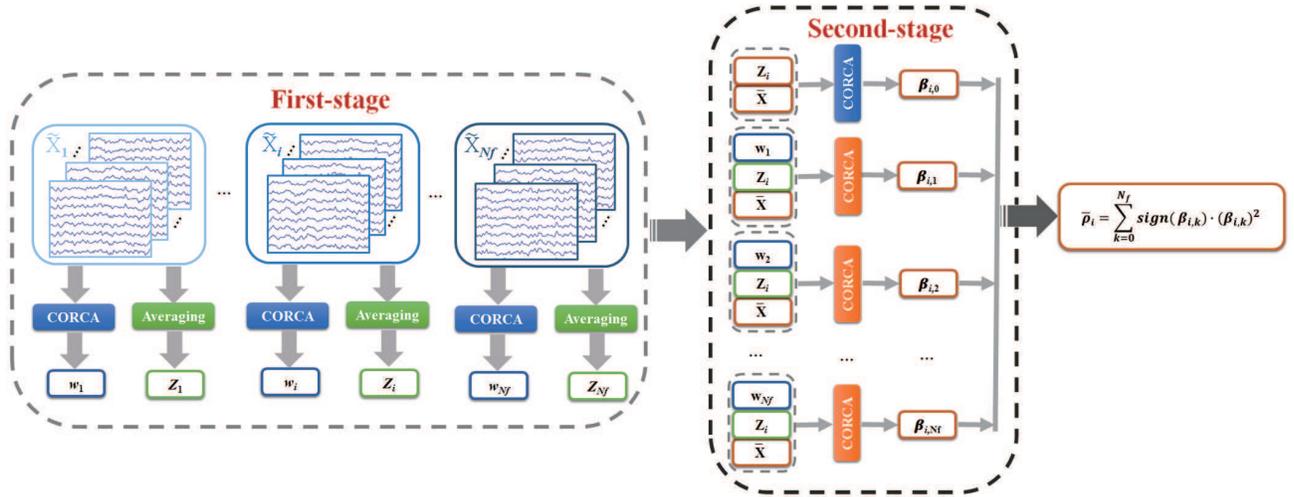}
\caption{Flowchart of the proposed two-stage CORRCA-based method. For each subject, $N_{f}$ training data corresponding to all the stimulus frequencies are available, $\bm {\widetilde X}_i \in \mathbb{R}^{Nc \times Ns \times Nt},i = 1, 2,\ldots,N_f$. In the first stage, the spatial filters for each frequency, i.e., $ w_1,w_2,\ldots,w_{N_f}$, are generated with formulas (6)-(8), and the reference signals are generated by group averaging across multiple training blocks, $ \bm{Z}_1,\bm{Z}_2,\ldots,\bm{Z}_{N_f}$. In the second stage, with a test sample $\bar{\bm{X}} \in \mathbb{R}^{C \times N}$  and an individual template $\bm{Z}_i \in \mathbb{R}^{C \times N} $ ($i=1,2,\ldots,N_f$), we can first calculate the correlation coefficient between $\bm{Z}_i$ and $\bar{\bm{X}}$ with formula (\ref{corrca2}), denoted as $\beta_{i,0}$. Then, with the weight vector $\bm{w}_k$, we can further calculate the correlation coefficients $\beta_{i,k}$ between $\bm{Z}_i$ and $\bar{\bm{X}}$ using formula (\ref{betacombi}). These correlation values are further combined as the feature by formula (\ref{rcombination}) .}
\label{fig:fig1}
\end{center}
\end{figure*}

The diagram of the proposed method is shown in Fig. \ref{fig:fig1}.

\subsection{The trial of filter bank technology}
Filter bank technology has been widely adopted in algorithm development for recent BCI systems \cite{Tanaka2017}. This technology could enhance the performance of original algorithms, such as the common spatial pattern (CSP) algorithm \cite{ang2012filter,zhangYu2015}, and CCA \cite{chen2015filter,Tanaka2017}. Therefore, we further investigate the results when a filter bank is added in the proposed methods, i.e., CORRCA and TSCORRCA. Here, five filter banks were used, and  the lower and upper cut-off frequencies of the $i$-th ($i=1,\cdots,5$) subband were set to $i\times8$ Hz and 90 Hz, respectively. The zero-phase Chebyshev Type I infinite impulse response (IIR) was used to extract each subband signal. The procedure for combining features in all subbands was similar to that in reference \cite{zhangcorca2018}.

\subsection{The exploration on cross-subject classification}
Exploiting the inter-subject information can reduce the training time \cite{yuan2015enhancing}. We evaluate the performance of standard CORRCA when the reference signals were transfered from the other existing subjects. We used the leave-one-out strategy to compute the reference signals for each subject. Concretely, for each subject, the data from the other thirty-four subjects in the benchmark dataset are used for reference signal computation, i.e., by group averaging. Here, standard CCA was used for comparison.

\section{Experimental Study}
\subsection{EEG recordings}
The data used in the current study were from an existing database, which was provided in the reference \cite{wangyj2016}. For the data collection, thirty-five healthy subjects (seventeen females, mean age 22 years) participated in an offline 40-target BCI speller experiment. The speller contains 40 stimuli coded at different frequencies (8-15.8 Hz with an interval of 0.2 Hz). For each subject, the experiment included six blocks. Each block contained 40 trials corresponding to all 40 stimuli indicated in a random order. Each trial lasted a total of 6 s, which consisted of  0.5 s for the visual cue and 0.5 s for stimulus offset before the next trial began. In each block, the subjects were asked to avoid eye blinks during the stimulation period. To avoid visual fatigue, there was a rest for several minutes between two consecutive blocks.

EEG data were recorded with a Synamps2 system (Neuroscan, Inc.) at a sampling rate of 1000 Hz  with a 0.15 Hz to 200 Hz bandpass filter and a notch filter at 50 Hz. All data were recorded from sixty-four channels that were placed on the standard positions according to the international 10-20 system.  The ground electrode (GND) was placed midway between Fz and FPz. The reference electrode was located on the vertex (Cz). Electrode impedances were maintained below 10 $k \Omega$. Event triggers were generated by the computer to the amplifier and were recorded on an event channel synchronized to the EEG data. The continuous EEG data were segmented into 6 s epochs (0.5 s prestimulus, 5.5 s poststimulus onset). The epochs were subsequently downsampled to 250 Hz. More detailed information for this dataset can be found in reference \cite{wangyj2016}.

\subsection{Performance evaluation}
In the current study, an extensive comparison was performed among the standard CCA method, the combination method of CCA and IT-CCA (named CCAICT), and the proposed standard CORRCA and two-stage CORRCA methods  (named TSCORRCA). A leave-one-out cross validation was employed to evaluate the classification accuracy of the four methods. Specifically, the EEG samples from five blocks were used for the training set, and the samples from the single left-out block were used for the testing set. The procedure was repeated six times such that each run was used as the testing set once.

For the CCA method, the recognition accuracy is directly evaluated by six runs of validation since no training process is required.

In this study, we also evaluated the feature values for each method using the $r$-square value, which was defined as the proportion of the variance of the signal feature that is accounted for by the user's intent \cite{wolpaw2002}. In the current study, the $r$-square value was calculated with the feature values of the target stimulus and the maximal feature values of the nontarget stimuli \cite{nakanishi2015}.

\section{Results}
\begin{figure}[htbp]
\begin{center}
\includegraphics[width=0.45\textwidth]{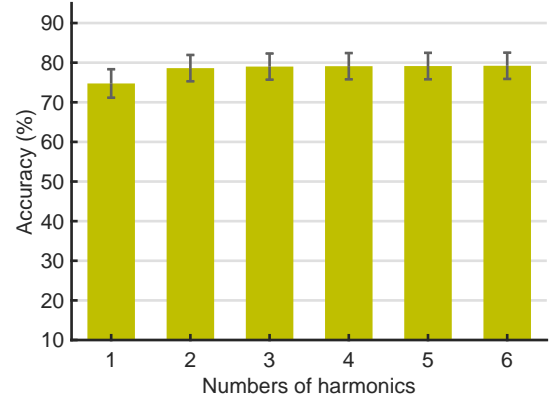}
\caption{Average classification accuracy for the CCAICT method with different numbers of harmonics. Here, TW is 0.8 s.}
\label{fig:fig2}
\end{center}
\end{figure}

Previous studies have indicated that the selection of the number of harmonics ($N_h$) plays an important role in the CCA method. Fig.~\ref{fig:fig2} shows the classification accuracy of CCAICT at different $N_h$ values in the reference signals in formula (\ref{rcorr}) with a data length of 0.8 s. Overall, the classification accuracy increased as the number of harmonics increased. One-way repeated-measures analysis of variance (ANOVA) showed that there were significant differences between different numbers of harmonics. Pairwise comparisons revealed significant differences between $N_h=1$ and all the other $N_h$ values. For a fair and convincing comparison, in the following computation, the number of harmonics was set to five as that in the reference \cite{chen2015PNAS}, which includes the fundamental and harmonic components of SSVEPs.

\begin{figure*}[htbp]
\begin{center}
\includegraphics[width=0.95\textwidth]{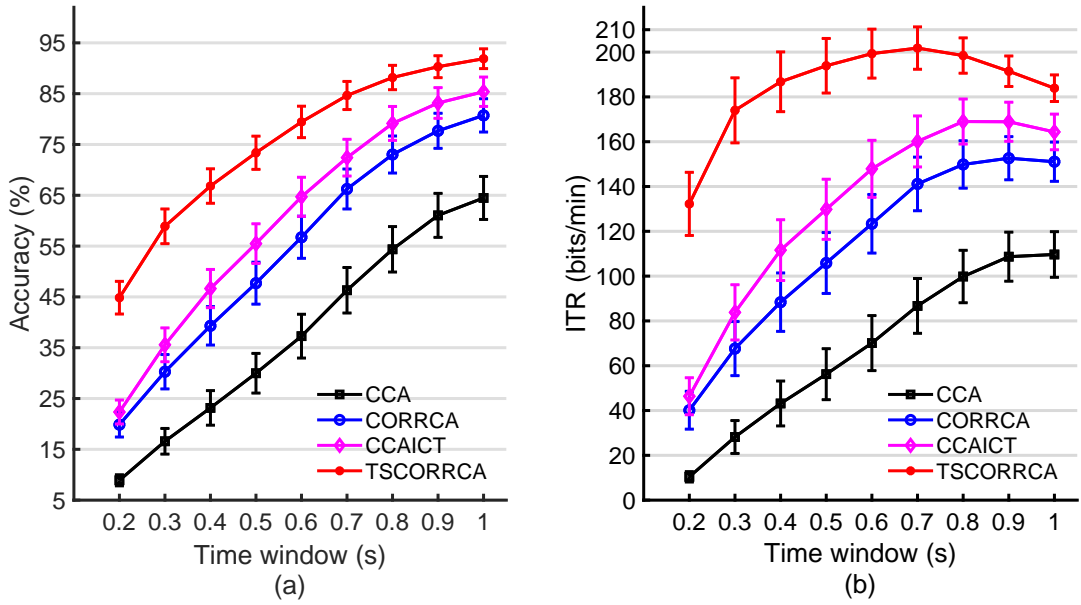}
\caption{ Average results across subjects of the four methods using different time windows. (a) Average classification accuracy and (b) simulated ITRs.  Error bars indicate standard errors.}
\label{fig:fig3}
\end{center}
\end{figure*}

Fig.~\ref{fig:fig3} shows the average accuracies and simulated ITRs across all subjects with different TWs. The standard CORRCA outperforms the standard CCA method, and TSCORRCA yields the best performance compared with all other methods. One-way repeated-measures ANOVA showed that there were significant differences in the classification accuracy between these methods at all TWs and significant differences in the simulated ITRs. The statistical analysis results are summarized in Table \ref{tab:table1}. Furthermore, for the accuracy and ITR, post hoc paired $t$-tests showed that there were significant differences between all pairs of the four methods at each TW ($p<0.001$).

\begin{table*}[htbp]
\renewcommand{\arraystretch}{1.2}
\scriptsize
\caption{ Statistical analysis of the classification accuracy differences between these methods at various time windows. $r$ denotes correlation coefficients, and $p$ denotes the significance level of the correlation coefficients. }
 \vspace*{-12pt}\label{T-Acc}
\begin{center}
\def\temptablewidth{\textwidth}
{\rule{\temptablewidth}{1pt}}
\begin{tabular*}{\temptablewidth}{@{\extracolsep{\fill}}ccccccccccc}
\multirow{2}{*}{ } & \multirow{2}{*}{ } & \multicolumn{9}{c}{Time windows} \\
\cline{3-11}
               &  &  $0.2 s$  &   $0.3 s$  &   $0.4 s$ &   $0.5 s$  &    $0.6 s$  &   $0.7 s$  &   $0.8 s$  &    $0.9 s$  &    $1 s$\\
\hline

\multirow{2}{*}{Accuracy} & F(3,102) & 39.51 & 30.63 & 25.44 & 22.25 & 20.55 & 18.22 & 16.30 & 13.96 & 13.40 \\
\cline{2-11}
                       &$p$ & $<$0.001 & $<$0.001 & $<$0.001 &$<$0.001 & $<$0.001 & $<$0.001  & $<$0.001 & $<$0.001 & $<$0.001\\
\hline
\multirow{2}{*}{ITR} & F(3,102) & 31.93 & 27.03 & 22.78 & 20.36 & 19.10 & 17.82 & 16.66 & 14.62 & 14.16 \\
\cline{2-11}
                        &$p$ & $<$0.001 & $<$0.001 & $<$0.001 &$<$0.001 & $<$0.001 & $<$0.001  & $<$0.001 & $<$0.001 & $<$0.001\\

\end{tabular*}
{\rule{\temptablewidth}{1pt}}
\end{center}
\label{tab:table1}
\end{table*}

\begin{figure*}[htbp]
\begin{center}
\includegraphics[width=0.95\textwidth]{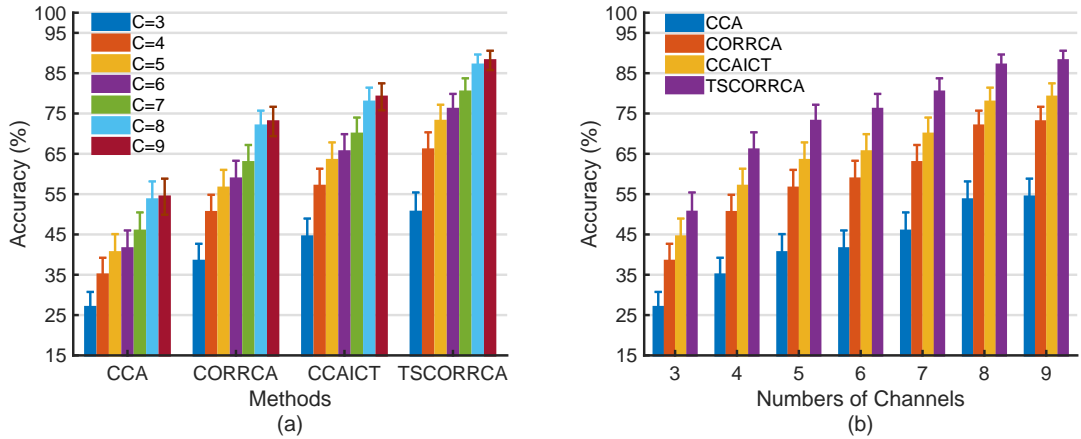}
\caption{ Average accuracy across subjects for each method using different numbers of channels. Error bars indicate standard errors. Here, TW is 0.8 s.}
\label{fig:fig4}
\end{center}
\end{figure*}

To further evaluate the performance among the four methods, we investigated the effects of different numbers of channels and training blocks on the classification accuracy. Fig.~\ref{fig:fig4}(a) shows the classification accuracy for each method with different numbers of channels at a 0.8 s TW. For all methods, the classification accuracy tended to increase with increasing number of channels. One-way repeated-measures ANOVA showed significant differences between different numbers of channels for all methods (CCA: $F(6,204)$=5.08, $p<0.001$; CORRCA: $F(6,204)$=8.46, $p<0.001$; CCAICT: $F(6,204)=8.86$, $p<0.001$; and TSCORRCA: $F(6,204)=12.61$, $p<0.001$). As shown in Fig.~\ref{fig:fig4}(b), TSCORRCA achieved the best performance, and CORRCA outperformed CCA at all numbers of channels. One-way repeated-measures ANOVA showed significant differences between the four methods at each condition ($C=3$: $F(3,102)=5.38$, $p=0.002$; $C=4$: $F(3,102)=9.41$, $p<0.001$; $C=5$: $F(3,102)=9.93$, $p<0.001$; $C=6$: $F(3,102)=11.72$, $p<0.001$; $C=7$: $F(3,102)=12.65$, $p<0.001$; $C=8$: $F(3,102)=15.09$, $p<0.001$; and $C=9$: $F(3,102)=16.30$, $p<0.001$).

\begin{figure*}[htbp]
\begin{center}
\includegraphics[width=0.95\textwidth]{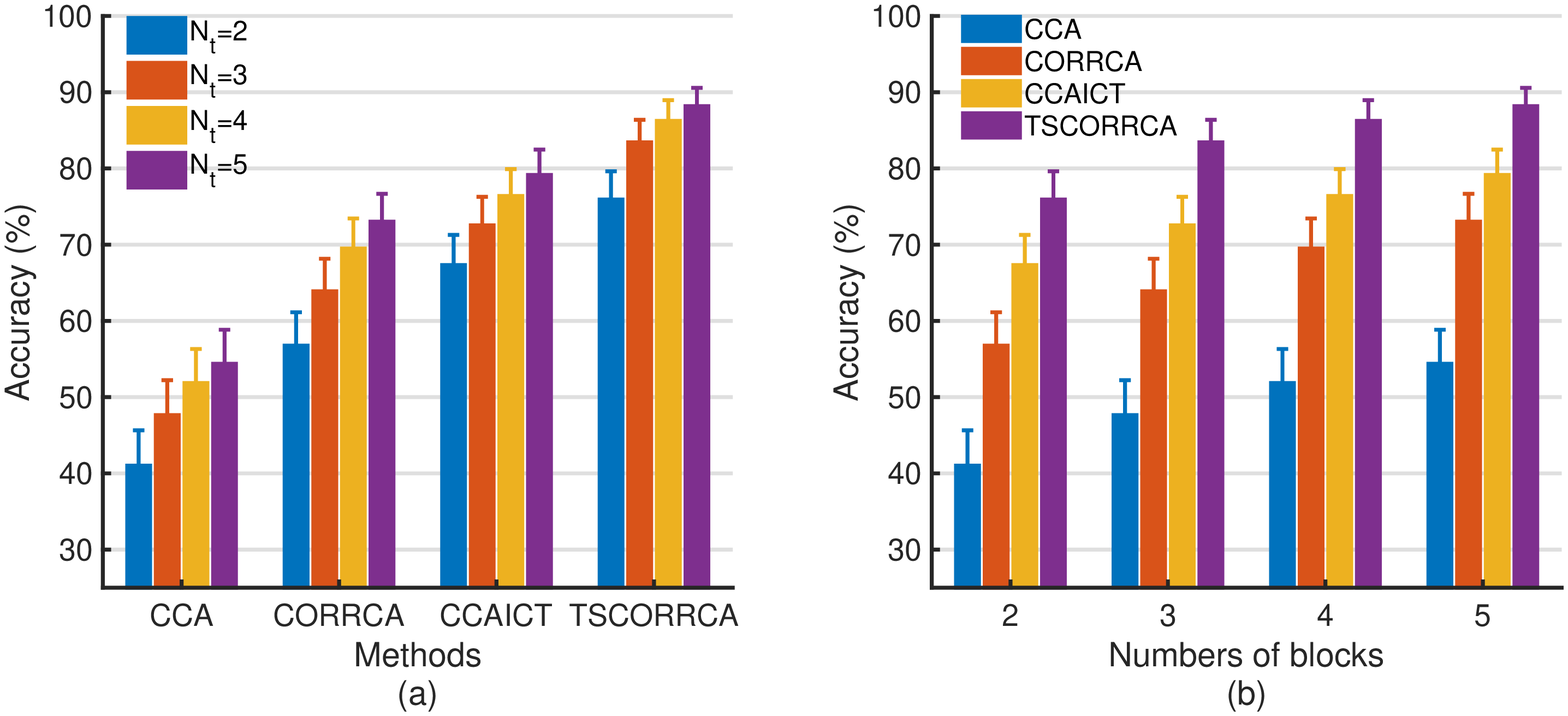}
\caption{Average accuracy across subjects with different numbers of training blocks for each method. Error bars indicate standard errors. Here, TW is 0.8 s.}
\label{fig:fig5}
\end{center}
\end{figure*}

Fig.~\ref{fig:fig5}(a) shows the classification accuracy for each method with different numbers of training blocks at a 0.8 s TW. Overall, the classification accuracy increased with increasing number of training blocks. However, one-way repeated-measures ANOVA showed that there were no significant differences between the numbers of training blocks for CCA ($F(3,102)=1.65$, $p=0.18$) and CCAICT ($F(3,102)=1.98$, $p=0.12$), but there were significant differences for CORRCA ($F(3,102)=3.05$, $p=0.03$) and TSCORRCA ($F(3,102)=3.25$, $p=0.02$). Furthermore, as shown in Fig.~\ref{fig:fig5}(b), TSCORRCA has the best performance among the methods, and the standard CORRCA outperformed the standard CCA at all numbers of training blocks. One-way repeated-measures ANOVA showed significant differences between the four methods at each condition ($N_t$=2: $F(3,102)=12.92$, $p<0.001$; $N_t$=3: $F(3,102)=14.75$, $p<0.001$; $N_t$=4: $F(3,102)=15.20$, $p<0.001$; and $N_t$=5: $F(3,102)=16.30$, $p<0.001$).

\begin{figure*}[htbp]
\begin{center}
\includegraphics[width=0.95\textwidth]{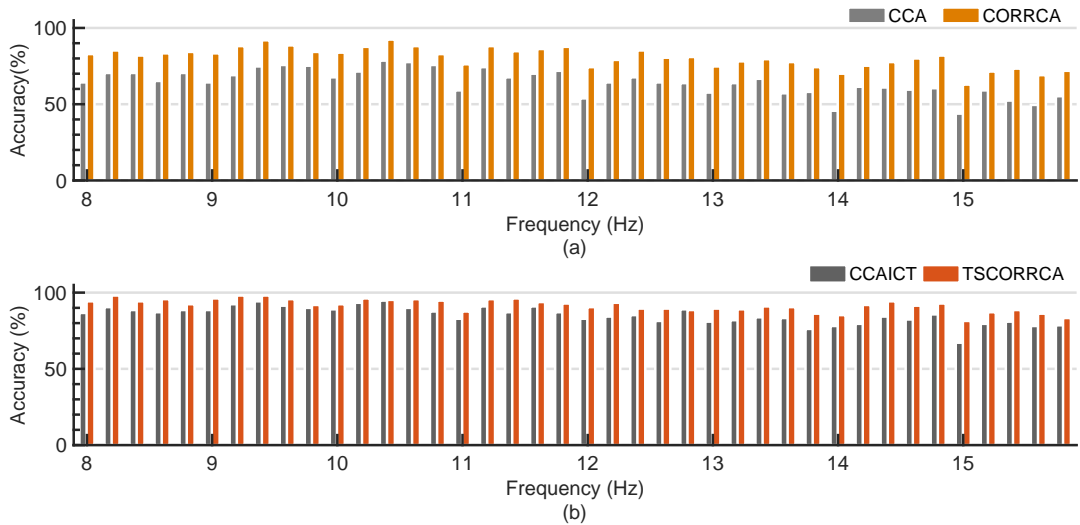}
\caption{Accuracy averaged across all subjects at each of the forty stimulus frequencies for the four methods at a 1 s time window.}
\label{fig:fig6}
\end{center}
\end{figure*}

\begin{figure}[p]
\begin{center}
\includegraphics[width=0.45\textwidth]{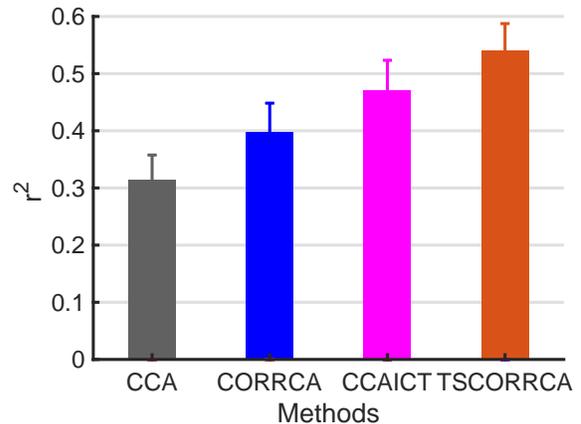}
\caption{$r$-square values for SSVEPs at 8.2 Hz. Error bars in each subfigure indicate standard errors. Here, TW is 0.8 s.}
\label{fig:fig7}
\end{center}
\end{figure}

In Fig.~\ref{fig:fig6}, we present the recognition accuracy averaged on all subjects at each of the forty stimulus frequencies for the four methods at a 1 s TW. CORRCA achieves better performance than CCA (Fig.~\ref{fig:fig6}(a)), and  TSCORRCA achieves overall better performance than CCAICT (Fig.~\ref{fig:fig6}(b)). To further explore the efficiency, $r$-square values obtained at 8.2 Hz are shown in Fig.~\ref{fig:fig7}. The TW was also set to 0.8 s. One-way repeated-measures ANOVA showed a significant difference between these methods ($F(3,102)=3.97$, $p=0.01$), and post hoc paired $t$-tests showed that there were significant differences between the combination method and the other methods. The results indicate that the proposed methods, i.e., CORRCA and TSCORRCA, can enhance the discriminability compared to CCA and CCAICT and then facilitate target classification.

\begin{figure}[htbp]
\begin{center}
\includegraphics[width=0.45\textwidth]{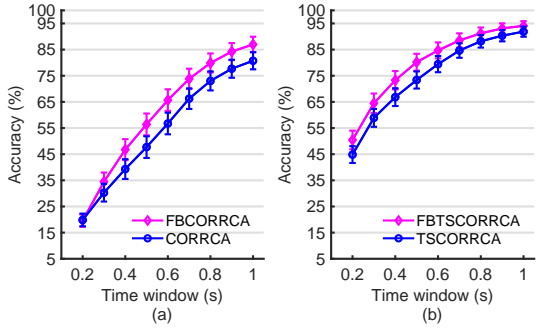}
\caption{The average accuracies across all subjects obtained by the CORRCA and TSCORRCA methods with a filter bank at various time windows. The error bars indicate standard errors. FBCORRCA and FBTSCORRCA denote the CORRCA and TSCORRCA methods with a filter bank, respectively.}
\label{fig:fig8}
\end{center}
\end{figure}

Filter bank technology could enhance the performance of algorithms in BCI systems. Here, we investigated the performance of the CORRCA and TSCORRCA with filter bank at the various TWs. As we expected, we found that the classification accuracies of both methods were improved with filter bank as shown in Fig.~\ref{fig:fig8}.

\begin{figure}[htbp]
\begin{center}
\includegraphics[width=0.45\textwidth]{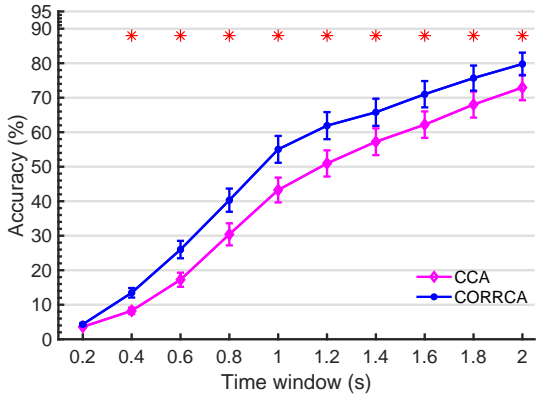}
\caption{The average accuracies across all subjects obtained by the standard CORRCA and CCA methods when the SSVEP reference signals were computed with  datasets from other subjects at various time windows. The error bars indicate standard errors. The asterisk indicates the statistically significant differences (paired $t$-test, $p<0.001$)}
\label{fig:fig9}
\end{center}
\end{figure}

For the results in Fig.~\ref{fig:fig3}, the test data and reference are acquired from the same subject. We further evaluated the performance of standard CORRCA when the reference signals were transfered from the other existing subjects. Fig.~\ref{fig:fig9} illustrates the average accuracies at various TWs using the standard CORRCA and CCA methods. As shown, the CORRCA still yields better performance than CCA, although the results are worse than those when the reference signals were obtained from the same subject. These findings demonstrate that CORRCA could be a promising method for designing and implementing a high-performance method for SSVEP frequency detection. Developing more efficient methods with CORRCA by exploiting intersubject information is beyond the scope of current paper, but we will work on this topic in future studies. 

\section{Discussion}
It is still a challenging issue to design and explore high-efficiency algorithms to classify EEG signals in BCI systems. Many algorithms have been proposed for different types of BCI modalities \cite{Lotte2018review}. For the SSVEP-based BCI, CCA is a state-of-the-art frequency recognition method, and it is widely used by the research community. To date, various extended methods have been developed \cite{nakanishi2015}, among which a combination method of CCA and IT-CCA achieved the best performance \cite{chen2015PNAS}. However, CCA requires the canonical vectors to be orthogonal, which may be not a reasonable assumption for EEG analysis. In fact, the spatial distributions are not expected to be orthogonal because they are determined by the current source distributions in space and the anatomy of the brain \cite{Dmochowski2012}. Moreover, the projection vectors for the two multichannel signals obtained by CCA are different. When two signals are generated by the same subjects, it is appropriate that the projection vectors should be the same. For instance, in the current study, the two multichannel signals, i.e., the test sample and the reference signals, were recorded from the same subject. Thus, the projection vectors should be the same. The vectors obtained by CCA were different, which may unnecessarily reduce the classification accuracy.

\begin{figure}[htbp]
\begin{center}
\includegraphics[width=0.45\textwidth]{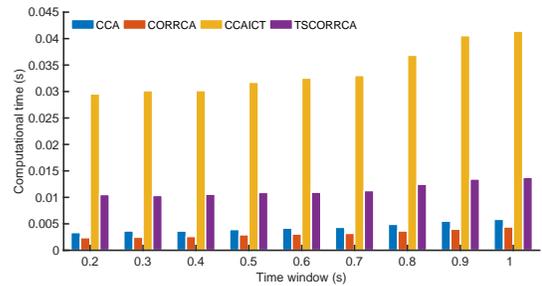}
\caption{The average computational complexities of the four methods at various time windows.}
\label{fig:fig10}
\end{center}
\end{figure}

In the current study, we proposed using CORRCA rather than CCA to implement frequency recognition. CORRCA could relax the constraint of canonical vectors in CCA and generate the same projection vector for two multichannel EEG signals. The experimental results show that the standard CORRCA method outperforms the standard CCA method when evaluated on the benchmark dataset and demonstrate the rationality and feasibility of CORRCA for SSVEP frequency recognition. We further extended the standard CORRCA method to a hierarchical method with two-stage operation, and the resulting  performance was significantly enhanced and better than that of the extended CCA-based method, i.e., IT-CCA. Compared with IT-CCA, the two-stage CORRCA method (TSCORRA) does not require the extra synthetic reference signals and thus does not need to optimize the number of harmonics ($N_h$). Furthermore, the computational efficiency was also compared among the four methods, and the results are shown in Fig.~\ref{fig:fig10}. The computational time was evaluated with MATLAB R2014b on a desktop computer with a 3.60 GHz CPU (16 GB RAM) at various TWs. We can find that all of the methods can be executed efficiently. Additionally, the CORRCA methods can be implemented faster than CCA, and TSCORRCA can be implemented faster than CCAICT.

CORRCA was firstly introduced for frequency detection  in our previous study \cite{zhangcorca2018}. In that study, we mainly used CORRCA to learn spatial filters with multiple blocks of individual training data. Then, the spatial filters were used to remove interference by combining the multichannel EEG signals. The feature extraction procedure needed extra one-dimensional or two-dimensional correlation analysis computation to obtain the features between the test sample and reference signals. Therefore, it was a different method compared to the methods proposed here. Overall, those methods demonstrate the feasibility and efficiency of CORRCA for frequency recognition. EEG signals are nonlinear and nonstationary. Thus far, we only considered the linear transformations in all the CORRCA-based methods. We will explore extending the methods to nonlinear versions with kernel methods \cite{parra2018correlated}, which may further improve the classification performance. In the current study, only the weight vectors corresponding to the maximum correlation coefficients were considered. In the future, we will investigate the performance of the methods with more weight vectors.

In recent years, some elaborately designed methods, such as deep-learning-based methods were developed for frequency detection \cite{KwakCNN,waytowich2018}. In the study \cite{KwakCNN}, the average classification rate in the static condition was $99.28\%$ on a 5-class SSVEP dataset. In another study \cite{waytowich2018}, the average accuracy was approximately $80\%$ on a 12-class SSVEP dataset. It seems that the proposed methods may not always exhibit better performance than these methods. However, we can find that the number of stimulus frequencies used in the two studies is much smaller than that used in our study. What's more, our proposed methods have low computational complexity, as shown in Fig.~\ref{fig:fig10}, and can easily be implemented. Accordingly, they may be good candidates for the BCI community to use in their BCI applications. It is appropriate and interesting to compare our methods with various deep-learning-based methods. Direct comparison of our method with those methods may be beyond the scope of this paper, we will endeavor on this topic in our future studies.

\section{Conclusion}
In summary, we proposed novel frequency recognition methods based on the CORRCA method. We confirmed that the standard CORRCA outperformed the standard state-of-the-art CCA method for frequency recognition with a large number of stimuli on a benchmark dataset. We further proposed a two-stage CORRCA method, which has the best performance compared to the most efficient method based on CCA. The experimental results suggest that the two-stage CORRCA method is a promising candidate to achieve satisfactory performance for SSVEP frequency recognition in BCI applications.

\section*{Acknowledgments}
The authors sincerely thank the reviewers for their insightful comments. The authors also sincerely thank Professor Lucas C. Parra and his colleagues generously share the correlated components analysis related codes. The authors acknowledge Dr. Dong Li for valuable comments on an early version of this manuscript.

\bibliographystyle{IEEEtran}



\end{document}